\documentclass[a4paper,11pt]{article}
\pdfoutput=1
\usepackage{jheppub}
\usepackage[T1]{fontenc}
\usepackage[utf8]{inputenc}
\usepackage{amsthm,mathtools}
\usepackage{booktabs,array}
\hypersetup{pdftitle={Liouville CFT, Matrix Models, and constrained WZW},pdfauthor={Babak Haghighat}}
\newtheorem{proposition}{Proposition}[section]

\newcommand{\slc}{\mathfrak{sl}_2}
\newcommand{\dd}{\mathrm{d}}

\newcommand{\Tr}{\operatorname{Tr}}
\newcommand{\cU}{\mathcal U}

\newcommand{\cS}{\mathcal S}

\allowdisplaybreaks
\title{\boldmath Liouville CFT, Matrix Models, and constrained WZW}
\author[\ast,\dag]{Babak Haghighat}
\affiliation[\ast]{Yau Mathematical Sciences Center, Tsinghua University, Beijing, 100084, China}
\affiliation[\dag]{Yanqi Lake Beijing Institute of Mathematical Sciences and Applications (BIMSA), Huairou District, Beijing 101408, P. R. China}
\abstract{
We study the interplay between screening integrals, $\beta$-deformed matrix models, and conformal blocks in Liouville and $\mathfrak{sl}_2$ Wess--Zumino--Witten (WZW) theories in the presence of irregular singularities. By framing the classical relation between the two theories through Drinfeld--Sokolov reduction, we show how the two-component first-order operator KZ system naturally transforms into a scalar second-order oper equation. Local solutions of this equation are constructed in terms of screening integrals, and their leading singular coefficients are matched directly with the semiclassical eigenvalues characterizing Liouville irregular states. Finally, we demonstrate how Wakimoto screening currents, together with the contravariant pairing on highest-weight modules, generate the lowering operators that govern integral representations of WZW conformal blocks.
}
\begin{document}
\maketitle

\section{Introduction}\label{sec:intro}
Two-dimensional Liouville conformal field theory (CFT) stands at a remarkable crossroads of modern mathematical physics, linking two- and three-dimensional quantum gravity~\cite{Saad:2019lba,Mertens:2020hbs,Collier:2023fwi,Collier:2024mgv,Collier:2024kmo}, four-dimensional supersymmetric gauge theories via the AGT correspondence~\cite{AGT}, and conformal matrix models~\cite{MarshakovMironovMorozov1991,KharchevEtAl1993,DijkgraafVafa,Kostov}. Conformal blocks provide a unifying language across these diverse areas: Virasoro symmetry, affine Kac--Moody symmetry, and free-field Coulomb-gas screening techniques offer complementary perspectives on closely related differential equations and contour integrals. This interplay becomes especially rich in the presence of irregular singularities, where the familiar highest-weight conditions are replaced by coherent-state eigenvalue conditions for positive Virasoro or affine current modes~\cite{Gaiotto,GaiottoTeschner}. Understanding how these structures translate into one another requires keeping track of the distinct mathematical roles played by scalar Coulomb-gas kernels, vector-valued affine conformal blocks, and the gauge or cohomological reductions that bridge them.

The goal of this article is twofold. First, in Part~I, we provide a self-contained, pedagogical review of the core ingredients connecting Liouville theory, $\beta$-deformed eigenvalue integrals, and chiral $\slc$ Wess--Zumino--Witten (WZW) theory. Second, in Part~II, we develop an explicit classical dictionary at the level of meromorphic connections and show how the standard affine screening mechanism incorporates representation-theoretic data beyond the reduced Liouville screening current. Specifically, we demonstrate how the classical Drinfeld--Sokolov gauge reduction converts the auxiliary first-order linear problem of the operator KZ equation into a second-order Sturm--Liouville oper, how the Miura transformation automatically cancels non-Cartan reconstruction fields via the stress-tensor improvement, and how Wakimoto $\beta\gamma$ contractions generate the spin-lowering operators of Schechtman--Varchenko integral solutions.

\subsection{Physical motivation and mathematical background}
The foundational framework for conformal blocks and degenerate Virasoro representations goes back to Belavin, Polyakov, and Zamolodchikov (BPZ)~\cite{BPZ}, where null-vector decoupling yields linear differential equations for correlation functions. Shortly thereafter, Dotsenko and Fateev~\cite{DF} showed that chiral blocks in the presence of degenerate insertions admit explicit contour-integral representations built from dimension-one free-field screening currents. While the full nonperturbative Liouville theory on the sphere requires the Dorn--Otto--Zamolodchikov--Zamolodchikov (DOZZ) three-point function and continuous bootstrap techniques~\cite{DO,ZZ,Teschner}, the Coulomb-gas integrals capture specific chiral sectors and their analytic continuations, providing a direct bridge to matrix-model eigenvalue integrals.

A parallel story unfolds in theories with affine Lie algebra symmetry, where current-algebra Ward identities and the Sugawara construction lead to the first-order Knizhnik--Zamolodchikov (KZ) system for vector-valued conformal blocks~\cite{KZ}. To construct free-field representations at generic level $k$, one passes from the level-one lattice vertex-operator construction of Frenkel and Kac~\cite{FrenkelKac} to the Wakimoto realization~\cite{Wakimoto}, which pairs a Heisenberg chiral boson with a first-order bosonic $\beta\gamma$ system. Within this framework, Schechtman and Varchenko~\cite{SchechtmanVarchenko} established that hypergeometric integrals over twisted cycles—weighting a scalar master function by rational, representation-valued functions—furnish solutions to the KZ equations, a structure deeply intertwined with the Gaudin model and the Bethe ansatz at the critical level~\cite{FeiginFrenkelReshetikhin}.

Two distinct yet complementary mechanisms relate $\slc$ WZW conformal blocks to Liouville theory. The first is Hamiltonian or Drinfeld--Sokolov (DS) reduction~\cite{DS}, which imposes a constraint on a nilpotent current and quotients by the residual gauge symmetry. In the two-dimensional lagrangian formulation, constraining opposite chiral currents in a Gauss decomposition reduces the WZW equations of motion directly to the classical Liouville equation~\cite{Forgacs,BershadskyOoguri}, while at the quantum level the reduction is implemented cohomologically via BRST ghosts~\cite{FeiginFrenkel}. A second, alternative bridge is the Ribault--Teschner correspondence~\cite{RibaultTeschner} and its path-integral derivation~\cite{HikidaSchomerus}, which relate $n$-point $H_3^+$ WZW correlators to Liouville correlators containing additional degenerate fields located at the zeros of a nilpotent current—a phenomenon naturally understood through the geometry of separation of variables~\cite{FrenkelGukovTeschner}.

The extension of these ideas to irregular singularities was sparked on the Virasoro side by the discovery that four-dimensional asymptotically free and Argyres--Douglas gauge theories correspond to confluence limits of primary vertex operators~\cite{Gaiotto,MarshakovMironovMorozov2009,GaiottoTeschner}. In the Coulomb-gas picture, these irregular states deform the logarithmic screening kernel by polynomial potentials, turning the conformal blocks into $\beta$-deformed matrix integrals governed by loop equations and Virasoro constraints~\cite{DijkgraafVafa,Kostov,ChekhovEynard,NishinakaRim,Rim,ChoiRimZhang}. On the affine side, irregular singularities have an equally rich life: confluent KZ equations and their free-field integral solutions govern isomonodromic deformations and quantum Painlev\'e equations~\cite{Reshetikhin,JimboNagoyaSun,NagoyaSun,Nagoya,NagoyaSunConfluent,NagoyaIrregular}, irregular Gaudin models describe opers with higher-order poles at the critical level~\cite{FFTL}, and singular Kac--Moody modules provide a natural representation-theoretic home for irregular WZW blocks and their flat connections~\cite{FelderRembado,GaiottoLamyPoirier,Haghighat,GukovHaghighatLiuReshetikhin}.

\subsection{Plan of the paper and summary of results}
In Part~I, we set up our free-field conventions and review the interplay between screening charges, matrix models, and WZW reduction. Section~\ref{sec:liouville} reviews chiral Liouville data, degenerate insertions, and the coherent-state realization of rank-$r$ irregular Virasoro states. Section~\ref{sec:ensembles} connects these screening integrals to analytically continued $\beta$-ensembles, deriving the exact finite-$N$ loop equations and showing how the collective eigenvalue field naturally incorporates the Liouville background charge. Section~\ref{sec:affine} introduces the Wakimoto realization and regular/irregular KZ equations, and Section~\ref{sec:relations} compares classical Hamiltonian reduction, quantum BRST reduction, and the separation-of-variables correspondence.

In Part~II, we turn to our main calculations. In Section~\ref{sec:classical}, we study the classical counterpart of the operator KZ equation—a two-component first-order linear system with meromorphic current coefficients—and perform its explicit gauge reduction to a scalar second-order oper equation. Specializing to the constrained Drinfeld--Sokolov slice with constant nilpotent current, we show that the derivative improvement in the oper potential automatically cancels the non-Cartan $\gamma$-field dependence, leaving the clean Miura combination $t=\alpha^2-\alpha'$ whose fundamental solutions are built from the classical screening primitive $S(z)=\int^z e^{2\phi(w)}\,\dd w$. In Section~\ref{sec:semiclassical}, we analyze the behavior of this dictionary at an irregular puncture and prove that, in a controlled large-$b$ semiclassical limit, the leading singular coefficients of $t(z)$ match the scaled Virasoro eigenvalues of the corresponding irregular coherent state. Finally, in Section~\ref{sec:integrals}, we bridge the classical and quantum pictures by showing explicitly how Wick contractions of the $\beta$ field in the unreduced Wakimoto screening current, combined with the contravariant bilinear form on highest-weight $\slc$ modules, generate the tensor-site lowering operators $B(w)=\sum_a F_a/(w-z_a)$ in integral solutions of the KZ equation.

\subsection{Conventions}
We work throughout with the complexified chiral algebra $\slc$ with generators satisfying
\begin{equation}\label{eq:levels}
 [H,E]=2E,\qquad [H,F]=-2F,\qquad [E,F]=H,\qquad \kappa=k+2\ne0,
\end{equation}
where $k$ is the affine level, $\kappa$ is the dual-Coxeter-shifted level, and $\ell_a=2j_a$ labels highest-weight representations (corresponding to finite-dimensional spin-$j_a$ representations when $\ell_a\in\mathbb Z_{\ge0}$). Our Liouville and matrix-ensemble parameters are related by
\begin{equation}\label{eq:parameters}
 Q=b+b^{-1},\qquad \kappa=-b^2,\qquad
 \beta_{\mathrm{ens}}=\kappa^{-1}=-b^{-2},
\end{equation}
where the last relation refers to Coulomb-gas integrals built from the $V_{1/b}$ screening current, and we write $\beta_{\mathrm{ens}}$ to distinguish the Dyson ensemble exponent from the Wakimoto field $\beta(z)$. We denote the quantum chiral Liouville boson by $\varphi(z)$ and its rescaled counterpart by $\phi(z)=\varphi(z)/b$, with Miura current $\alpha(z)=\partial\phi(z)$. In Part~II, the same symbols $\phi$ and $\alpha$ will also denote their commuting semiclassical limits, while the full two-dimensional Liouville field is written as $\Phi_L$ and external Liouville momenta are distinguished by context.

\newpage
\part{Review of conformal blocks, screenings and reduction}

\section{Liouville theory and irregular Virasoro states}\label{sec:liouville}
\subsection{Chiral data and screening operators}
In Euclidean signature on a Riemann surface $\Sigma$ with metric $g_{ab}$ and scalar curvature $R$, the Liouville action takes the standard form
\begin{equation}
 S_L=\frac{1}{4\pi}\int_\Sigma\sqrt{g}\,
 \bigl(g^{ab}\partial_a\Phi_L\partial_b\Phi_L+Q R\Phi_L
       +4\pi\mu_L e^{2b\Phi_L}\bigr)\,\dd^2x,
\end{equation}
where $b$ is the Liouville coupling constant, $Q=b+b^{-1}$ is the background charge, and $\mu_L$ is the cosmological constant. In the chiral free-field (or Coulomb-gas) approach, we work with a holomorphic boson $\varphi(z)$ normalized such that
\begin{equation}\label{eq:liouvillefree}
 \varphi(z)\varphi(w)\sim-\frac12\log(z-w),\qquad
 T_L=-:(\partial\varphi)^2:+Q\partial^2\varphi.
\end{equation}
This stress tensor generates a Virasoro algebra with central charge $c_L=1+6Q^2$, under which primary vertex operators carry conformal weight
\begin{equation}
 V_\alpha=:e^{2\alpha\varphi}:,\qquad
 \Delta_\alpha=\alpha(Q-\alpha).
\end{equation}
The two screening currents $V_b$ and $V_{1/b}$ both have conformal dimension one, so their OPE with $T_L(z)$ is a total derivative in the screening coordinate. Consequently, integrating either current over a closed twisted cycle (or a relative cycle on which boundary terms vanish) yields a screening charge that commutes with the Virasoro algebra.

Degenerate representations of the Virasoro algebra are labeled by two positive integers $r,s\in\mathbb Z_{\ge1}$ and correspond to the Kac momenta
\begin{equation}
 \alpha_{r,s}=\frac{1-r}{2}b+\frac{1-s}{2b},\qquad
 V_{-\ell/(2b)}=V_{\alpha_{1,\ell+1}},\qquad \ell\in\mathbb Z_{\ge0}.
\end{equation}
While the vertex operator $V_{-\ell/(2b)}$ is well-defined for arbitrary complex $\ell$, at nonnegative integer $\ell$ the corresponding Verma module develops a singular vector at level $\ell+1$. Passing to the irreducible quotient decouples this null vector and imposes a BPZ differential equation of order $\ell+1$ on correlation functions~\cite{BPZ}, with the fundamental degenerate field $V_{\alpha_{1,2}}=V_{-1/(2b)}$ yielding a second-order equation.

To construct a Coulomb-gas conformal block with finite insertions $V_{-\ell_a/(2b)}(z_a)$ and $N$ screening currents $V_{1/b}(w_i)$, one evaluates the free-field Wick contractions to obtain the multivalued integrand
\begin{equation}\label{eq:liouvillekernel}
 \prod_{a<b}(z_a-z_b)^{-\ell_a\ell_b/(2b^2)}
 \prod_{i,a}(w_i-z_a)^{\ell_a/b^2}
 \prod_{i<j}(w_i-w_j)^{-2/b^2}.
\end{equation}
On the sphere, charge neutrality against a state of momentum $\alpha_\infty$ at infinity imposes the selection rule
\begin{equation}
 \alpha_\infty-\sum_a\frac{\ell_a}{2b}+\frac{N}{b}=Q.
\end{equation}
Selecting a specific basis of twisted integration cycles $\Gamma$ for the screening coordinates $w_i$ then furnishes a basis of chiral conformal blocks, which can subsequently be analytically continued in the external momenta beyond integer screening numbers.

\subsection{Irregular states as parameterized families}\label{sec:irregularreview}
Irregular Virasoro states arise when primary operators collide or when positive Virasoro generators act as coherent-state eigenvalues rather than annihilating the state~\cite{Gaiotto,GaiottoTeschner}. In the free-field realization, expanding the $U(1)$ current as $j(z)=\partial\varphi(z)=\sum_n j_nz^{-n-1}$ gives the Heisenberg algebra
\begin{equation}
 [j_n,j_m]=-\frac n2\delta_{n+m,0}.
\end{equation}
Starting from a Fock highest-weight vector $|c_0\rangle$ with $j_0|c_0\rangle=-c_0|c_0\rangle$, we can construct a rank-$r$ coherent family parameterized by $\mathbf c=(c_0,c_1,\dots,c_r)$ via
\begin{equation}\label{eq:coherent}
 |\mathbf c\rangle=
 \exp\left(\sum_{s=1}^r\frac{2c_s}{s}j_{-s}\right)|c_0\rangle.
\end{equation}
By construction, this state satisfies the coherent eigenvalue relations $j_s|\mathbf c\rangle=-c_s|\mathbf c\rangle$ for $1\le s\le r$ and is annihilated by higher modes $j_s$ for $s>r$. Expressing the Virasoro generators from \eqref{eq:liouvillefree} in normal-ordered form (with creation modes $j_{-s}$ placed to the left of annihilation modes $j_s$),
\begin{equation}
 L_n=-\sum_p:j_{n-p}j_p:-(n+1)Qj_n,
\end{equation}
one finds that the nonnegative Virasoro modes act on the family $|\mathbf c\rangle$ as first-order differential operators in the irregular parameters (see Appendix~\ref{app:modes} for details):
\begin{align}\label{eq:irregularward}
 L_n|\mathbf c\rangle
 &=\left(\Lambda_n+\sum_{s=1}^{r-n}s c_{n+s}\partial_{c_s}\right)|\mathbf c\rangle,\\
 \Lambda_n&=(n+1)Q c_n-\sum_{i=0}^n c_i c_{n-i},\qquad c_i=0\quad(i>r),\label{eq:lambdan}
\end{align}
where empty sums are understood to vanish. In particular, the modes in the window $r\le n\le2r$ act purely by the scalar eigenvalues $\Lambda_n$, modes with $n>2r$ annihilate $|\mathbf c\rangle$, and the dilatation generator acts as
\begin{equation}
 L_0|\mathbf c\rangle=
 \left(c_0(Q-c_0)+\sum_{s=1}^r s c_s\partial_{c_s}\right)|\mathbf c\rangle.
\end{equation}
Note that because \eqref{eq:irregularward} acts on a parameterized ket, successive Virasoro operators compose in reverse order on the parameter space—that is, $L_mL_n|\mathbf c\rangle=D_nD_m|\mathbf c\rangle$ where $D_n$ denotes the differential operator in \eqref{eq:irregularward}—which ensures consistency with the Virasoro commutation relations.

Equations \eqref{eq:irregularward}--\eqref{eq:lambdan} provide a transparent free-field realization of the Gaiotto--Teschner irregular module conditions~\cite{Gaiotto,GaiottoTeschner}. In constructing complete irregular conformal blocks, this local Fock-space state is supplemented by screening contours and appropriate asymptotic normalizations. When the irregular puncture is placed at $z=\infty$, one works in the local coordinate $\zeta=1/z$, where polynomial potentials in the finite screening coordinates $w_i$ naturally arise from singular terms at $\zeta=0$ after accounting for the conformal transformation of $j(z)$ and $T_L(z)$ (see, e.g., Sections~3.4--3.5 of~\cite{GukovHaghighatLiuReshetikhin} for explicit rank-one irregular integral formulas).

\section{Eigenvalue integrals and Virasoro constraints}\label{sec:ensembles}
\subsection{Formal beta ensembles}
The Coulomb-gas integrals for irregular Liouville conformal blocks naturally take the form of $\beta$-deformed matrix-model eigenvalue integrals~\cite{DijkgraafVafa,ItoyamaMaruyoshiOota2010,MironovMorozovShakirov2010,SchiappaWyllard2010,NishinakaRim,Rim,ChoiRimZhang}, analytically continued to complex coupling parameters and integration cycles:
\begin{equation}\label{eq:ensemble}
 Z_N(\mathbf t)=\frac1{N!}\int_\Gamma
 \prod_{i=1}^N\dd w_i\,
 \Delta(\mathbf w)^{2\beta_{\mathrm{ens}}}
 \exp\left(\sum_{i=1}^N U(w_i)\right),\qquad
 U(w)=\sum_{s\ge0}t_s w^s,
\end{equation}
where $\Delta(\mathbf w)=\prod_{i<j}(w_i-w_j)$ is the Vandermonde determinant. While the classical Dyson indices $2\beta_{\mathrm{ens}}=1,2,4$ correspond to ensembles of real symmetric, Hermitian, and quaternionic self-dual random matrices, the Liouville identification \eqref{eq:parameters} sets $\beta_{\mathrm{ens}}=-b^{-2}$, so we view \eqref{eq:ensemble} as a formal contour integral over a twisted cycle $\Gamma$ in $\mathbb C^N$.

Comparing with the Coulomb-gas kernel \eqref{eq:liouvillekernel}, each finite degenerate field $V_{-\ell_a/(2b)}(z_a)$ shifts the one-body potential $U(w)$ by a logarithmic term $(\ell_a/b^2)\log(w-z_a)$, while a rank-$r$ irregular state at infinity introduces a polynomial potential of degree $r$ in $w$. Up to the $w$-independent prefactor $\prod_{a<b}(z_a-z_b)^{-\ell_a\ell_b/(2b^2)}$, the chiral Liouville conformal block is therefore directly captured by the eigenvalue partition function \eqref{eq:ensemble}.

\subsection{Resolvents, loop equations and time-variable constraints}\label{sec:loopsdetail}
We now derive the Ward identities of the eigenvalue integral, following the collective-field approach of~\cite{Kostov} while keeping the $\beta$-deformation explicit. For brevity, we write $\beta$ for $\beta_{\mathrm{ens}}$ throughout this subsection and keep the exponential weight $e^{\sum_i U(w_i)}$ of \eqref{eq:ensemble} (which is related to the customary Hermitian-matrix potential $e^{-\Tr V(M)}$ by $U=-V$).

\paragraph{From matrices to eigenvalue observables.}
At the Hermitian point $\beta=1$, diagonalizing an $N\times N$ Hermitian matrix $M$ and integrating out the unitary angular degrees of freedom yields
\begin{equation}
 \int\dd M\,e^{\Tr U(M)}
 =\frac1{N!}\int\prod_{i=1}^N\dd w_i\,
 \Delta(\mathbf w)^2 e^{\sum_i U(w_i)},
\end{equation}
where $\Delta(\mathbf w)^2$ is the Jacobian of diagonalization and $1/N!$ mods out by Weyl-group permutations of the eigenvalues. Replacing $\Delta(\mathbf w)^2$ by $\Delta(\mathbf w)^{2\beta}$ defines the $\beta$-deformed integral \eqref{eq:ensemble} directly in eigenvalue space for arbitrary complex $\beta$.

For any symmetric observable $\mathcal O(\mathbf w)$, its normalized expectation value is defined by
\begin{equation}\label{eq:ensembleaverage}
 \langle\mathcal O\rangle=
 \frac{1}{N!Z_N}\int_\Gamma\prod_i\dd w_i\,
 \mathcal O(\mathbf w)\Delta(\mathbf w)^{2\beta}
 e^{\sum_i U(w_i)}.
\end{equation}
The fundamental collective observable of the model is the resolvent
\begin{equation}\label{eq:resolvent}
 R(z)=\sum_{i=1}^N\frac1{z-w_i},\qquad
 R(z)=\Tr\frac1{z-M}\quad\text{in the matrix realization},
\end{equation}
which acts as the generating function for the eigenvalue power sums $p_s=\sum_i w_i^s$ (with $p_0=N$) via its formal large-$z$ expansion
\begin{equation}\label{eq:resolventexpansion}
 R(z)=\sum_{s\ge0}\frac{p_s}{z^{s+1}}.
\end{equation}
When the integration contour lies along the real axis, one can also introduce the empirical eigenvalue density $\widehat\rho(x)=\sum_i\delta(x-w_i)$, in terms of which the one-point and connected two-point resolvent correlators read
\begin{equation}
 \langle R(z)\rangle=\int\frac{\langle\widehat\rho(x)\rangle}{z-x}\,\dd x,
 \qquad
 \langle R(z)R(\zeta)\rangle_c
 =\int\frac{\langle\widehat\rho(x)\widehat\rho(y)\rangle_c}
 {(z-x)(\zeta-y)}\,\dd x\,\dd y,
\end{equation}
where the subscript $c$ denotes the connected cumulant and $\widehat\rho(x)\widehat\rho(y)$ includes the contact terms at coincident eigenvalues $i=j$. On general complex screening cycles $\Gamma$, we work directly with the algebraic definition \eqref{eq:resolvent}.

Because the coupling constants $t_s$ in $U(w)=\sum_{s\ge0}t_s w^s$ act as sources for the power sums $p_s$, differentiating the partition function at fixed integration cycle $\Gamma$ inserts power sums into the integrand:
\begin{equation}\label{eq:powersumderivatives}
 \partial_{t_s}Z_N=Z_N\langle p_s\rangle,\qquad
 \partial_{t_s}\partial_{t_q}Z_N=Z_N\langle p_s p_q\rangle,
 \qquad \partial_{t_0}Z_N=N Z_N,
\end{equation}
while derivatives of the free energy $\log Z_N$ generate connected correlators,
\begin{equation}
 \partial_{t_s}\partial_{t_q}\log Z_N
 =\langle p_s p_q\rangle-\langle p_s\rangle\langle p_q\rangle.
\end{equation}
Introducing the loop-insertion operator $\mathcal D(z)=\sum_{s\ge0}z^{-s-1}\partial_{t_s}$ therefore inserts the resolvent $R(z)$ when acting on $Z_N$, and generates connected multi-resolvent correlators when acting on $\log Z_N$:
\begin{equation}
 \mathcal D(z)\mathcal D(\zeta)\log Z_N=\langle R(z)R(\zeta)\rangle_c.
\end{equation}

\paragraph{The integration-by-parts identity.}
Writing $\mathcal W(\mathbf w)=\Delta(\mathbf w)^{2\beta}e^{\sum_i U(w_i)}$ for the full integrand weight, its logarithmic derivative with respect to $w_i$ is
\begin{equation}
 \partial_{w_i}\log\mathcal W
 =U'(w_i)+2\beta\sum_{j\ne i}\frac1{w_i-w_j}.
\end{equation}
For $z$ away from the integration contour $\Gamma$, invariance of the integral under the infinitesimal reparameterization $\delta w_i=\epsilon/(z-w_i)$ on a closed or decaying twisted cycle yields the Schwinger--Dyson identity
\begin{align}\label{eq:ibpresolvent}
 0&=\frac1{N!Z_N}\sum_i\int_\Gamma\dd^Nw\,
 \partial_{w_i}\left(\frac{\mathcal W}{z-w_i}\right)\nonumber\\
 &=\left\langle\sum_i\frac1{(z-w_i)^2}
 +2\beta\sum_{i\ne j}\frac1{(z-w_i)(w_i-w_j)}
 +\sum_i\frac{U'(w_i)}{z-w_i}\right\rangle.
\end{align}
Symmetrizing the double sum in \eqref{eq:ibpresolvent} over pairs $(i,j)$ and $(j,i)$ gives
\begin{align}
 \sum_{i\ne j}\frac1{(z-w_i)(w_i-w_j)}
 &=\sum_{i<j}\frac1{w_i-w_j}
 \left(\frac1{z-w_i}-\frac1{z-w_j}\right)\nonumber\\
 &=\sum_{i<j}\frac1{(z-w_i)(z-w_j)}.
\end{align}
Combining this off-diagonal sum with the diagonal terms in $R(z)^2$, we obtain the algebraic identity
\begin{equation}
 2\sum_{i\ne j}\frac1{(z-w_i)(w_i-w_j)}
 =R(z)^2-\sum_i\frac1{(z-w_i)^2}.
\end{equation}
Noting that $\partial_zR(z)=-\sum_i(z-w_i)^{-2}$, the Schwinger--Dyson identity \eqref{eq:ibpresolvent} takes the compact loop-equation form
\begin{equation}\label{eq:loop}
 \boxed{\left\langle
 \beta R(z)^2-(1-\beta)\partial_zR(z)
 +\sum_i\frac{U'(w_i)}{z-w_i}
 \right\rangle=0.}
\end{equation}
While the derivative term $\partial_z R(z)$ drops out at the Hermitian-matrix point $\beta=1$, at generic $\beta$ it plays a crucial role: it directly generates the Liouville background-charge improvement in the collective CFT stress tensor.

To separate the singular and polynomial parts in $z$, we introduce
\begin{equation}\label{eq:polynomialremainder}
 f(z)=\sum_i\frac{U'(z)-U'(w_i)}{z-w_i},
\end{equation}
which is a polynomial in $z$ whenever $U(w)$ is a polynomial. The loop equation \eqref{eq:loop} can then be rewritten as
\begin{equation}\label{eq:loopremainder}
 \beta\langle R(z)^2\rangle-(1-\beta)\partial_z\langle R(z)\rangle
 +U'(z)\langle R(z)\rangle-\langle f(z)\rangle=0,
\end{equation}
where at finite $N$ the quadratic expectation value decomposes into disconnected and connected pieces as $\langle R(z)^2\rangle=\langle R(z)\rangle^2+\langle R(z)R(z)\rangle_c$.

\paragraph{From loop equations to Virasoro constraints.}
Expanding \eqref{eq:ibpresolvent} in inverse powers of $z$ at infinity corresponds to performing the polynomial coordinate variations $\delta w_i=\epsilon w_i^{n+1}$ for $n\ge-1$. At mode number $n\ge0$, we find
\begin{equation}
 0=\left\langle(n+1)p_n
 +2\beta\sum_{i\ne j}\frac{w_i^{n+1}}{w_i-w_j}
 +\sum_{s\ge1}s t_s p_{s+n}\right\rangle.
\end{equation}
Symmetrizing the Vandermonde interaction term once more yields
\begin{align}
 2\sum_{i\ne j}\frac{w_i^{n+1}}{w_i-w_j}
 &=\sum_{i\ne j}\frac{w_i^{n+1}-w_j^{n+1}}{w_i-w_j}\nonumber\\
 &=\sum_{p=0}^{n}\sum_{i\ne j}w_i^p w_j^{n-p}
 =\sum_{p=0}^{n}p_p p_{n-p}-(n+1)p_n.
\end{align}
Adding the measure Jacobian and the pair interaction leaves the relation
\begin{equation}
 \left\langle\beta\sum_{p=0}^{n}p_p p_{n-p}
 +(1-\beta)(n+1)p_n+\sum_{s\ge1}s t_s p_{s+n}\right\rangle=0.
\end{equation}
By virtue of the source identities \eqref{eq:powersumderivatives}, these expectation values translate into linear differential constraints on the partition function,
\begin{equation}\label{eq:timesvir}
 \mathcal L_n Z_N=0,\qquad n\ge-1,
\end{equation}
governed by the $\beta$-deformed Virasoro operators
\begin{equation}\label{eq:Lnensemble}
 \mathcal L_n=
 \beta\sum_{p=0}^n\partial_{t_p}\partial_{t_{n-p}}
 +\sum_{s\ge1}s t_s\partial_{t_{s+n}}
 +(1-\beta)(n+1)\partial_{t_n},\qquad n\ge0,
\end{equation}
together with the $n=-1$ translation generator
\begin{equation}
 \mathcal L_{-1}=\sum_{s\ge1}s t_s\partial_{t_{s-1}}.
\end{equation}
In particular, the lowest modes $n=-1$ and $n=0$ express overall translation and scale invariance of the eigenvalue integral,
\begin{equation}
 \sum_{s\ge1}s t_s\langle p_{s-1}\rangle=0,\qquad
 \beta N^2+(1-\beta)N+\sum_{s\ge1}s t_s\langle p_s\rangle=0,
\end{equation}
where the $N$-dependent terms in the dilatation identity reflect the scaling weight $N$ of the integration measure $\prod_i \dd w_i$ plus the weight $\beta N(N-1)$ of the Vandermonde factor $\Delta(\mathbf w)^{2\beta}$. When logarithmic insertions $\sum_a (\ell_a/b^2)\log(w-z_a)$ are also included in $U(w)$, the fundamental loop equation \eqref{eq:loop} continues to hold, with the rational poles at $z=z_a$ encoding derivatives with respect to the insertion points $z_a$.

\subsection{The collective field and its free-boson realization}\label{sec:collective}
The structure of the Virasoro operators \eqref{eq:Lnensemble} naturally invites us to package the coupling constants and time derivatives into a single chiral collective field. Let us first introduce this field as an eigenvalue observable inside the integral and then pass to its Fock-space representation on the space of coupling parameters.

\paragraph{A collective field built from the eigenvalues.}
Choosing a branch of $\sqrt{2\beta}$ (with $\beta\ne0$) and writing $U_+(z)=\sum_{s\ge1}t_s z^s$, we define the collective chiral boson and its associated $U(1)$ current by
\begin{equation}\label{eq:collectivefield}
 \Phi_{\mathrm{coll}}(z)
 =\frac{U_+(z)}{\sqrt{2\beta}}
 +\sqrt{2\beta}\sum_i\log(z-w_i),\qquad
 J_{\mathrm{coll}}(z)=\partial_z\Phi_{\mathrm{coll}}(z)
 =\frac{U'(z)}{\sqrt{2\beta}}+\sqrt{2\beta}R(z).
\end{equation}
In the matrix realization, the logarithmic sum is simply $\Tr\log(z-M)$, whose derivative is the resolvent $R(z)$ and whose large-$z$ expansion reads
\begin{equation}\label{eq:logexpansion}
 \sum_i\log(z-w_i)=N\log z-\sum_{s\ge1}\frac{p_s}{s z^s},
\end{equation}
where the zero mode $N\log z$ is separated explicitly from the power-sum modes $s\ge1$.

Setting $\rho=(1-\beta)/\sqrt{2\beta}$, we form the improved collective energy--momentum tensor
\begin{equation}\label{eq:Tcollective}
 T_{\mathrm{coll}}(z)=\frac12J_{\mathrm{coll}}(z)^2
 -\rho\,\partial_zJ_{\mathrm{coll}}(z).
\end{equation}
Substituting \eqref{eq:collectivefield} yields the explicit expansion
\begin{equation}\label{eq:Tcollectiveexpanded}
 T_{\mathrm{coll}}=
 \beta R^2+U'R-(1-\beta)R'
 +\frac{U'^2}{4\beta}-\frac{1-\beta}{2\beta}U''.
\end{equation}
Notice how the coefficient $\rho$ of the background-charge improvement $-\rho\,\partial_z J_{\mathrm{coll}}$ is uniquely dictated by matching the derivative term $-(1-\beta)R'(z)$ in the loop equation \eqref{eq:loop}.

Denoting by $[\cdot]_{<0}$ the projection onto strictly negative powers of $z$ in a Laurent expansion at infinity, and observing that $f(z)$ in \eqref{eq:polynomialremainder}, $U'(z)^2$, and $U''(z)$ are all polynomials in $z$, the loop equation \eqref{eq:loopremainder} is equivalent to the compact statement
\begin{equation}\label{eq:collectiveprojection}
 [\langle T_{\mathrm{coll}}(z)\rangle]_{<0}=0.
\end{equation}
Since $[U'(z)R(z)]_{<0}=\sum_iU'(w_i)/(z-w_i)$, projecting onto the polar part $z^{-n-2}$ ($n\ge-1$) of $\langle T_{\mathrm{coll}}(z)\rangle$ reproduces the entire hierarchy of loop equations, while leaving the non-singular polynomial part of $\langle T_{\mathrm{coll}}(z)\rangle$ unconstrained.

\paragraph{Insertion operators and Heisenberg modes.}
Using the source relations \eqref{eq:powersumderivatives}, we can promote the collective field \eqref{eq:collectivefield} to a differential operator acting on the coupling space $\{t_s\}$:
\begin{equation}\label{eq:timeboson}
 \widehat\Phi(z)=\frac{U_+(z)}{\sqrt{2\beta}}
 +\sqrt{2\beta}\log z\,\partial_{t_0}
 -\sqrt{2\beta}\sum_{s\ge1}\frac{z^{-s}}{s}\partial_{t_s}.
\end{equation}
Differentiating with respect to $z$ gives the Heisenberg current operator
\begin{equation}\label{eq:timecurrent}
 \widehat J(z)=\sum_{n\in\mathbb Z}a_n z^{-n-1}
 =\frac{U'(z)}{\sqrt{2\beta}}
 +\sqrt{2\beta}\sum_{s\ge0}z^{-s-1}\partial_{t_s},
\end{equation}
whose modes are realized on functions of the coupling times by
\begin{equation}\label{eq:ensembleheisenberg}
 a_n=\sqrt{2\beta}\,\partial_{t_n},\qquad
 a_{-n}=\frac{n t_n}{\sqrt{2\beta}}\quad(n>0),\qquad
 a_0=\sqrt{2\beta}\,\partial_{t_0}.
\end{equation}
From $[\partial_{t_n},t_m]=\delta_{nm}$, one immediately verifies the canonical Heisenberg algebra
\begin{equation}
 [a_n,a_m]=n\delta_{n+m,0},\qquad
 \widehat J(z)\widehat J(w)\sim\frac1{(z-w)^2},
\end{equation}
where the zero mode $a_0$ measures the total eigenvalue number $\sqrt{2\beta}N$ on $Z_N$, the multiplication operators $a_{-n}$ ($n>0$) act as creation modes, and the derivatives $a_n$ ($n>0$) act as annihilation modes.

With normal ordering $:~:$ defined by placing creation operators $a_{-n}$ to the left of annihilation operators $a_n$, the normal-ordered stress tensor reads
\begin{equation}\label{eq:ensembleT}
 \widehat T(z)=\frac12:\widehat J(z)^2:
 -\rho\,\partial_z\widehat J(z)
 =\sum_{n\in\mathbb Z}\mathcal L_n z^{-n-2},
\end{equation}
with Virasoro generators
\begin{equation}\label{eq:ensemblevirasoromodes}
 \mathcal L_n=\frac12\sum_{p\in\mathbb Z}:a_p a_{n-p}:
 +\rho(n+1)a_n.
\end{equation}
For $n\ge0$, the annihilation--annihilation terms reproduce the second-order derivative piece,
\begin{equation}
 \frac12\sum_{p=0}^{n}a_p a_{n-p}
 =\beta\sum_{p=0}^{n}\partial_{t_p}\partial_{t_{n-p}},
\end{equation}
the mixed creation--annihilation terms yield
\begin{equation}
 \sum_{s\ge1}a_{-s}a_{n+s}
 =\sum_{s\ge1}s t_s\partial_{t_{n+s}},
\end{equation}
and the linear improvement term gives $\rho(n+1)a_n=(1-\beta)(n+1)\partial_{t_n}$, while for $n=-1$ only the mixed sum $\sum_{s\ge1}s t_s\partial_{t_{s-1}}$ survives. Thus \eqref{eq:ensemblevirasoromodes} reproduces the exact differential operators \eqref{eq:Lnensemble}. Moreover, because normal ordering prevents the derivative part of one current from differentiating the explicit coupling times $t_s$ inside the other current, acting with $\widehat T(z)$ on $Z_N$ directly inserts the collective stress tensor into the integral:
\begin{equation}\label{eq:operatorcollective}
 \widehat T(z)Z_N
 =Z_N\langle T_{\mathrm{coll}}(z)\rangle,
 \qquad
 [\widehat T(z)Z_N]_{<0}
 =\sum_{n\ge-1}z^{-n-2}\mathcal L_nZ_N=0.
\end{equation}

\paragraph{Background charge and the Liouville normalization.}
The operators $\mathcal L_n$ generate a Virasoro algebra with central charge determined by the background charge $\rho$:
\begin{equation}
 [\mathcal L_n,\mathcal L_m]
 =(n-m)\mathcal L_{n+m}
 +\frac{c}{12}(n^3-n)\delta_{n+m,0},\qquad
 c=1-12\rho^2.
\end{equation}
Indeed, in the OPE of $\widehat T(z)\widehat T(w)$, the double Wick contraction of $\tfrac12:\widehat J^2:$ contributes $1/[2(z-w)^4]$ and the contraction of $-\rho\,\partial\widehat J$ with itself contributes $-6\rho^2/(z-w)^4$, summing to $c/[2(z-w)^4]$ with
\begin{equation}\label{eq:ensemblecentral}
 c=1-6\frac{(1-\beta)^2}{\beta}.
\end{equation}
At $\beta=1$ this reduces to the unimproved $c=1$ free boson, whereas at the Liouville value $\beta=-b^{-2}$ (with branch choice $\sqrt{\beta}=i/b$) we obtain
\begin{equation}
 \rho=-\frac{i}{\sqrt2}(b+b^{-1})=-\frac{iQ}{\sqrt2}.
\end{equation}
Identifying the currents via $\widehat J=-i\sqrt2\,\partial\varphi$ matches our Liouville conventions \eqref{eq:liouvillefree} and turns \eqref{eq:ensembleT} into
\begin{equation}
 \widehat T=-:(\partial\varphi)^2:+Q\partial^2\varphi,
 \qquad c=1+6Q^2.
\end{equation}

\paragraph{Vertex insertions and passing to affine symmetry.}
In this eigenvalue picture, inserting a finite vertex operator of Liouville momentum $-\ell/(2b)$ introduces a power of the characteristic polynomial into the integrand,
\begin{equation}
 \mathcal O_\ell(z)=\prod_i(w_i-z)^{\ell/b^2},\qquad
 U(w)\longmapsto U(w)+\frac{\ell}{b^2}\log(w-z),
\end{equation}
whose logarithmic derivative with respect to $z$ generates $(\ell/b^2)R(z)$. At nonnegative integer $\ell$, this insertion corresponds to the degenerate field $V_{-\ell/(2b)}(z)$. While the scalar collective boson $\widehat\Phi(z)$ and its Virasoro constraints $\mathcal L_n Z_N=0$ ($n\ge-1$) fully govern the scalar Coulomb-gas kernel, passing to vector-valued WZW conformal blocks and the Knizhnik--Zamolodchikov equations at generic level $\kappa$ requires enriching this free boson with the first-order $\beta\gamma$ system of the Wakimoto realization~\cite{Wakimoto}, to which we now turn.

\section{Affine free fields, screenings and KZ equations}\label{sec:affine}
\subsection{Wakimoto conventions}
To realize the affine $\slc$ current algebra at generic level $k$ ($\kappa=k+2\ne0$), we work with the rescaled chiral boson $\phi=\varphi/b$ and its $U(1)$ current $\alpha=\partial\phi$, supplemented by a bosonic first-order $(\beta,\gamma)$ system of conformal weights $(1,0)$. Their fundamental operator product expansions are
\begin{align}\label{eq:wakOPE}
 \phi(z)\phi(w)&\sim\frac{1}{2\kappa}\log(z-w),&
 \alpha(z)\alpha(w)&\sim\frac{1}{2\kappa(z-w)^2},\\
 \beta(z)\gamma(w)&\sim\frac1{z-w},&
 \gamma(z)\beta(w)&\sim-\frac1{z-w}.
\end{align}
In terms of these free fields, the Wakimoto realization~\cite{Wakimoto} of the $\slc$ currents takes the form
\begin{align}\label{eq:wakimoto}
 J^+&=\beta,\qquad J^0=-2\kappa\alpha-2:\gamma\beta:,\\
 J^-&=-:\gamma^2\beta:-2\kappa:\gamma\alpha:+k\partial\gamma,
\end{align}
which obey the standard affine $\slc$ OPEs at level $k$:
\begin{equation}
 J^0(z)J^0(w)\sim\frac{2k}{(z-w)^2},\qquad
 J^+(z)J^-(w)\sim\frac{k}{(z-w)^2}+\frac{J^0(w)}{z-w}.
\end{equation}
The associated Sugawara energy--momentum tensor decomposes into a $\beta\gamma$ piece of central charge $2$ and a bosonic piece with background charge:
\begin{equation}\label{eq:sugawara}
 T_{\mathrm{Sug}}=:\beta\partial\gamma:
 +\kappa:\alpha^2:+\partial\alpha,\qquad
 c_{\mathrm{Sug}}=\frac{3k}{k+2}.
\end{equation}
With respect to $T_{\mathrm{Sug}}$, a purely bosonic vertex operator $:e^{q\phi}:$ carries conformal dimension $q(q-2)/(4\kappa)$.

\subsection{Affine and Liouville screening currents}
The Wakimoto screening current is constructed by dressing the bosonic exponential $V(w)=:e^{2\phi(w)}:$ with the weight-one field $\beta(w)$:
\begin{equation}\label{eq:affscreen}
 \cS(w)=:\beta(w)e^{2\phi(w)}:.
\end{equation}
This composite current has conformal weight one and regular OPEs with $J^+(z)$ and $J^0(z)$, while its OPE with the lowering current $J^-(z)$ is a total derivative with respect to $w$:
\begin{equation}\label{eq:screenOPE}
 J^-(z)\cS(w)\sim
 \frac{\kappa V(w)}{(z-w)^2}
 +\frac{2\kappa:\alpha(w)V(w):}{z-w}
 =\kappa\partial_w\left(\frac{V(w)}{z-w}\right).
\end{equation}
Integrating $\cS(w)$ over a closed twisted cycle (or a relative cycle with vanishing boundary contributions) therefore yields a screening operator that intertwines the affine $\slc$ action.

Notice the essential role played by the $\beta(w)$ prefactor in \eqref{eq:affscreen}: when $\cS(w)$ is inserted into a free-field correlator alongside primary vertex operators, the bosonic exponential $:e^{2\phi(w)}:$ produces the scalar master function, whereas the Wick contractions of $\beta(w)$ with the $\gamma$-dependent primary fields generate the rational, tensor-product-valued weight functions of the Schechtman--Varchenko integral representation~\cite{SchechtmanVarchenko}. Upon passing to the Drinfeld--Sokolov reduced theory where $\beta$ is constrained to a constant, $\cS(w)$ reduces directly to the scalar Liouville screening current $V_{1/b}(w)=:e^{2\phi(w)}:$.

\subsection{Conformal blocks and regular KZ equations}
Chiral conformal blocks in WZW theory are defined as linear functionals on the tensor product of representations inserted at punctures $z_a$ that are invariant under the meromorphic current algebra—or, dually, as elements of the space of coinvariants. Choosing a basis in each representation represents the conformal block as a vector-valued function $\Psi(z_1,\dots,z_n)$. At noncritical level $\kappa\ne0$, the Sugawara identity $L_{-1}=\frac{1}{2\kappa}\sum_{m\in\mathbb Z}:J^a_{-1-m}J_{a,m}:$ translates coordinate derivatives into the Knizhnik--Zamolodchikov (KZ) system~\cite{KZ}:
\begin{equation}\label{eq:KZregular}
 \kappa\partial_{z_a}\Psi=
 \sum_{b\ne a}\frac{\Omega_{ab}}{z_a-z_b}\Psi,\qquad
 \Omega_{ab}=\frac12H_aH_b+E_aF_b+F_aE_b,
\end{equation}
where $E_a,F_a,H_a$ act on the $a$-th tensor factor (or by their transposes when components are assembled by evaluating a functional on basis vectors).

The hypergeometric integral construction of Schechtman and Varchenko~\cite{SchechtmanVarchenko,FeiginFrenkelReshetikhin} provides explicit solutions to \eqref{eq:KZregular} by integrating the product of the scalar master function and the rational $\slc$-weight function over twisted cycles. In Section~\ref{sec:integrals}, we will show explicitly how this rational weight function and its spin-lowering operators emerge from the $\beta\gamma$ Wick contractions of the Wakimoto screening current \eqref{eq:affscreen}.

\subsection{Irregular affine data}
Just as in the Virasoro case, one can introduce irregular punctures in WZW theory where a finite number of positive affine current modes act nontrivially on a cyclic state while higher positive modes annihilate it. Such states are naturally organized as induced representations of truncated current algebras (or singular Kac--Moody modules)~\cite{JimboNagoyaSun,NagoyaSun,NagoyaSunConfluent,FelderRembado,Haghighat,GukovHaghighatLiuReshetikhin}, with the Lie brackets of the positive modes constraining which currents can simultaneously acquire non-vanishing eigenvalues. (In a complementary direction at the critical level $\kappa=0$, irregular Gaudin models and opers with higher-order poles are constructed using the Feigin--Frenkel center~\cite{FFTL}.)

The simplest irregular deformation is a rank-one Cartan-type irregular singularity at infinity, which modifies the regular KZ system \eqref{eq:KZregular} by a constant Cartan term:
\begin{equation}\label{eq:irrKZ}
 \kappa\partial_{z_a}\Psi=
 \left(\frac{\Lambda}{2}H_a+
 \sum_{b\ne a}\frac{\Omega_{ab}}{z_a-z_b}\right)\Psi,
\end{equation}
defining a flat connection in the puncture coordinates $z_a$ at fixed irregular parameter $\Lambda$. Explicit integral solutions for this confluent KZ system and its higher-rank generalizations can be constructed via confluent Wakimoto screenings~\cite{NagoyaSun,GukovHaghighatLiuReshetikhin}.

\section{Bridges between WZW and Liouville theories}\label{sec:relations}
\subsection{Classical Hamiltonian reduction}
Before turning to our explicit connection-level dictionary in Part~II, it is illuminating to compare the three main ways in which $\slc$ WZW theory and Liouville theory are related.

At the classical level, Drinfeld--Sokolov (DS) reduction~\cite{DS} constrains a nilpotent component of the current and quotients by the residual unipotent gauge group. Choosing a canonical gauge slice brings the connection to Sturm--Liouville oper form, whose single non-trivial entry transforms as a projective connection (stress tensor), whereas choosing a triangular Miura representative expresses this potential as a quadratic differential in the Cartan current $\alpha$ plus a derivative improvement.

In the two-dimensional lagrangian WZW model~\cite{Forgacs}, one performs a local Gauss decomposition $g=ABC$ with upper- and lower-triangular factors parameterized by $a,\eta$ and a diagonal Cartan field $\phi_{2d}$. Imposing the chiral current constraints
\begin{equation}\label{eq:2dconstraints}
 \partial_+\eta=\mu e^{2\phi_{2d}},\qquad
 \partial_-a=\nu e^{2\phi_{2d}},
\end{equation}
reduces the WZW equations of motion directly to the two-dimensional Liouville equation for $\phi_{2d}$. Writing $a=-2\gamma$, the second relation in \eqref{eq:2dconstraints} can be integrated along $z_-$ to reconstruct the Gauss coordinate:
\begin{equation}\label{eq:gamma2d}
 \gamma(z_+,z_-)=\gamma_0(z_+)
 -\frac\nu2\int_{z_{-,0}}^{z_-}e^{2\phi_{2d}(z_+,s)}\,\dd s.
\end{equation}
Note that \eqref{eq:gamma2d} reconstructs the anti-holomorphic derivative $\partial_-\gamma$ of the two-dimensional Gauss coordinate in terms of the full field $\phi_{2d}(z_+,z_-)$, whereas the chiral oper reduction we will perform in Section~\ref{sec:classical} works purely holomorphically with the chiral Miura current $\alpha(z)=\partial\phi(z)$.

\subsection{Quantum DS reduction}
At the quantum level, Drinfeld--Sokolov reduction is formulated via BRST cohomology~\cite{FeiginFrenkel,BershadskyOoguri}. To make the constrained current $J^+(z)=\beta(z)$ a dimension-zero field so that the constraint $J^+(z)=1$ is conformally invariant, one twists the Sugawara stress tensor \eqref{eq:sugawara} by $\tfrac12\partial J^0$:
\begin{equation}
 T_{\mathrm{Sug}}+\frac12\partial J^0=
 -:\gamma\partial\beta:
 +\kappa:\alpha^2:
 +(1-\kappa)\partial\alpha.
\end{equation}
Upon introducing the fermionic $(b_{\mathrm{gh}},c_{\mathrm{gh}})$ ghost system for the nilpotent gauge symmetry and passing to BRST cohomology, the $\beta\gamma$ and ghost contributions cancel, leaving the reduced Virasoro stress tensor
\begin{equation}\label{eq:DSquantum}
 T_{\mathrm{DS}}=
 \kappa:\alpha^2:
 +(1-\kappa)\partial\alpha,
 \qquad c_{\mathrm{DS}}=13-6\kappa-\frac6\kappa.
\end{equation}
Under our parameter identification $\varphi=b\phi$ and $\kappa=-b^2$ from \eqref{eq:parameters}, $T_{\mathrm{DS}}$ matches the Liouville stress tensor \eqref{eq:liouvillefree} with $c_{\mathrm{DS}}=1+6Q^2$, while the bosonic exponential of the Wakimoto screening current \eqref{eq:affscreen} becomes the Liouville screening current:
\begin{equation}
 :e^{2\phi}:=:e^{2\varphi/b}:.
\end{equation}
Moreover, the improvement term $\tfrac12\partial J^0$ shifts the conformal dimensions of primary vertex operators: a bosonic highest-weight factor $:e^{-\ell\phi}:$, which has Sugawara dimension $\ell(\ell+2)/(4\kappa)$, acquires the shifted dimension
\begin{equation}
 \frac{\ell(\ell+2)}{4\kappa}-\frac\ell2
 =\Delta_{-\ell/(2b)},
\end{equation}
matching the exact conformal weight of the degenerate Liouville field $V_{-\ell/(2b)}$. At the level of representations, quantum DS reduction thus maps an affine module $M$ (including irregular modules) to its BRST cohomology $H^\bullet_{\mathrm{DS}}(M)$, which naturally carries a representation of the reduced Virasoro algebra \eqref{eq:DSquantum}.

\subsection{Separation of variables and correlation-function identities}
A third, complementary bridge between the two theories is the $H_3^+$ WZW--Liouville correspondence of Ribault and Teschner~\cite{RibaultTeschner} (see also~\cite{HikidaSchomerus} for a path-integral derivation and~\cite{GaiottoLamyPoirier} for the extension to irregular singularities). On the sphere, this correspondence relates an $n$-point WZW correlation function directly to a $(2n-2)$-point Liouville correlation function containing $n$ primary fields with shifted momenta together with $n-2$ degenerate fields $V_{-1/(2b)}(y_k)$ inserted at the zeros $y_k$ of the nilpotent current. Geometrically, as elucidated in~\cite{FrenkelGukovTeschner}, these extra degenerate punctures arise from passing to separated variables, where the zeros of the nilpotent current entry $e(z)$ create apparent singularities in the associated scalar oper equation.

\part{Explicit classical dictionary and affine screening contractions}

\section{Classical reduction of the operator-KZ linear problem}\label{sec:classical}
\subsection{The commuting-current problem}
At the quantum level, the operator KZ equation~\cite{GukovHaghighatLiuReshetikhin} governs the holomorphic derivative of a primary WZW field inserted in the fundamental representation through normal-ordered products with the affine $\slc$ currents. Passing to the classical limit where the currents become commuting meromorphic functions $(h,e,f)$, this relation takes the form of a local two-component first-order linear system:
\begin{equation}\label{eq:classicalsystem}
 \kappa\partial_z\Psi=A(z)\Psi,\qquad
 A=\begin{pmatrix}h&f\\ e&-h\end{pmatrix},
\end{equation}
where $h$ represents half the Cartan current in the fundamental representation and we keep the level parameter $\kappa$ explicit to track the connection normalization.

\begin{proposition}\label{prop:oper}
On any simply connected open patch where $e(z)$ is non-vanishing, choose a holomorphic branch of $e^{1/2}$ and define the triangular and diagonal gauge matrices
\begin{equation}\label{eq:gauges}
 D=\begin{pmatrix}e^{-1/2}&0\\0&e^{1/2}\end{pmatrix},\qquad
 H=h+\frac\kappa2\frac{e'}e,\qquad
 N=\begin{pmatrix}1&H\\0&1\end{pmatrix}.
\end{equation}
Then the gauge transformation $\Psi=DN\widehat\Psi$ brings the first-order system \eqref{eq:classicalsystem} into the canonical companion (oper) form
\begin{equation}\label{eq:oper}
 \kappa\partial_z\widehat\Psi=
 \begin{pmatrix}0&\cU\\1&0\end{pmatrix}\widehat\Psi,\qquad
 \cU=ef+H^2-\kappa H',
\end{equation}
whose lower component $\chi=\widehat\Psi_2$ satisfies the second-order scalar Sturm--Liouville equation $(\kappa^2\partial_z^2-\cU)\chi=0$.
\end{proposition}
\begin{proof}
Under a general gauge transformation $\Psi=G\widetilde\Psi$, the connection matrix transforms as $A\mapsto G^{-1}AG-\kappa G^{-1}G'$. Applying the diagonal transformation $D$ normalizes the lower-left entry to unity and shifts the diagonal entry to $H$, yielding $\bigl(\begin{smallmatrix}H&ef\\1&-H\end{smallmatrix}\bigr)$. Subsequent conjugation by the upper-triangular matrix $N$ eliminates the diagonal terms and produces \eqref{eq:oper}. Reading off the lower row gives $\kappa\chi'=\widehat\Psi_1$, and substituting this into the upper row yields $(\kappa^2\partial_z^2-\cU)\chi=0$.
\end{proof}
Expanding $\cU$ explicitly in terms of the original connection entries $(h,e,f)$ gives
\begin{equation}\label{eq:Uexpanded}
 \cU=h^2+ef-\kappa h'
 +\kappa h\frac{e'}e
 -\frac{\kappa^2}{2}\frac{e''}e
 +\frac{3\kappa^2}{4}\left(\frac{e'}e\right)^2.
\end{equation}
Geometrically, this gauge transformation implements the local Drinfeld--Sokolov reduction to an $\slc$-oper, where the lower component $\chi$ transforms globally as a $-1/2$-differential (half-density) and the zeros of $e(z)$ mark the points where the diagonal rescaling $D$ becomes singular.

\subsection{Constant nilpotent current and cancellation}\label{sec:cancellation}
Let us now specialize to the Drinfeld--Sokolov constrained slice where the nilpotent current is fixed to a non-zero constant, $e(z)=\mu\ne0$, and parameterize the remaining commuting currents in the classical Wakimoto form:
\begin{equation}\label{eq:classicalparam}
 h=-2\mu\gamma+\kappa\alpha,\qquad
 f=-4\mu\gamma^2+4\kappa\alpha\gamma-2\kappa\gamma'.
\end{equation}

\begin{proposition}\label{prop:cancellation}
Under the classical Wakimoto parameterization \eqref{eq:classicalparam} on the slice $e=\mu$, the reduced oper potential $\cU$ is completely independent of $\gamma(z)$ and takes the Miura form
\begin{equation}\label{eq:miura}
 t:=\frac{\cU}{\kappa^2}=\alpha^2-\alpha'.
\end{equation}
\end{proposition}
\begin{proof}
Substituting \eqref{eq:classicalparam} with $e=\mu$ into \eqref{eq:Uexpanded}, the quadratic Sugawara combination and the derivative improvement evaluate to
\begin{equation}
 h^2+ef=\kappa^2\alpha^2-2\kappa\mu\gamma',\qquad
 -\kappa h'=2\kappa\mu\gamma'-\kappa^2\alpha'.
\end{equation}
Adding the two expressions cancels the $-2\kappa\mu\gamma'$ terms identically, leaving $\cU=\kappa^2(\alpha^2-\alpha')$.
\end{proof}
This exact cancellation is a key feature of the reduction: while the unimproved quadratic combination $h^2+ef$ retains a term $-2\kappa\mu\gamma'$ proportional to the derivative of the non-Cartan field $\gamma$, the Drinfeld--Sokolov derivative improvement $-\kappa h'$ removes $\gamma'$ automatically, leaving a purely bosonic Miura potential $t=\alpha^2-\alpha'$ without any need to constrain $\gamma'$ separately.

\subsection{Solutions of the scalar equation from screening integrals}\label{sec:primitives}
Writing $\alpha=\phi'$ in terms of a classical chiral boson $\phi(z)$, the second-order oper operator factorizes into first-order factors:
\begin{equation}
 \partial_z^2-t=(\partial_z-\alpha)(\partial_z+\alpha).
\end{equation}
Consequently, a local fundamental basis of solutions to $(\partial_z^2-t)\chi=0$ is given explicitly in terms of the classical screening primitive $S(z)$ by
\begin{equation}\label{eq:solutions}
 \chi_1=e^{-\phi},\qquad
 \chi_2=e^{-\phi}S,\qquad
 S(z)=\int_{z_0}^{z}e^{2\phi(w)}\,\dd w.
\end{equation}
These two solutions have unit Wronskian, $\chi_1\chi_2'-\chi_1'\chi_2=1$, so their ratio $S=\chi_2/\chi_1$ defines the projective coordinate of the oper, whose Schwarzian derivative recovers the Miura potential:
\begin{equation}\label{eq:schwarzian}
 \{S,z\}:=\frac{S'''}{S'}-\frac32\left(\frac{S''}{S'}\right)^2=-2t.
\end{equation}
Varying the basepoint $z_0$ or choosing a different linear combination of $\chi_1,\chi_2$ acts on $S(z)$ by a M\"obius transformation.

The same screening integral $S(z)$ also appears naturally when solving the first-order matrix system \eqref{eq:classicalsystem} via a Gauss-type triangular factorization. Factoring out the upper-triangular $\gamma$-dependence via
\begin{equation}
 \Psi=\begin{pmatrix}1&-2\gamma\\0&1\end{pmatrix}\Psi_-,
\end{equation}
the parameterization \eqref{eq:classicalparam} reduces \eqref{eq:classicalsystem} to the lower-triangular system
\begin{equation}\label{eq:triangular}
 \partial_z\Psi_-=
 \begin{pmatrix}\alpha&0\\\mu/\kappa&-\alpha\end{pmatrix}\Psi_-.
\end{equation}
Writing a fundamental matrix solution in Gauss-decomposed form,
\begin{equation}
 \Psi_-=\begin{pmatrix}e^\phi&0\\0&e^{-\phi}\end{pmatrix}
 \begin{pmatrix}1&0\\\eta&1\end{pmatrix},\qquad
 \eta'=\frac\mu\kappa e^{2\phi},
\end{equation}
we immediately obtain $\eta(z)=\eta_0+(\mu/\kappa)S(z)$. Thus, in the holomorphic connection problem, the integral $S(z)=\int_{z_0}^z e^{2\phi(w)}\,\dd w$ directly reconstructs the lower-triangular Gauss coordinate $\eta(z)$. Since $\phi=\varphi/b$, the integrand $e^{2\phi}=e^{2\varphi/b}$ is precisely the classical counterpart of the Liouville screening current $V_{1/b}(w)=:e^{2\varphi(w)/b}:$.

\subsection{Zeros of the nilpotent entry}
It is instructive to examine what happens to the reduced oper potential \eqref{eq:Uexpanded} near a point $z=q$ where $h(z)$ and $f(z)$ are holomorphic while the nilpotent current $e(z)$ has a simple zero. Expanding \eqref{eq:Uexpanded} around $z=q$ yields a double pole of fixed residue:
\begin{equation}
 \cU(z)=\frac{3\kappa^2}{4(z-q)^2}
 +\frac{u_{-1}}{z-q}+O(1),
\end{equation}
whose indicial roots are $-1/2$ and $3/2$. Because the underlying first-order matrix connection $A(z)$ is completely regular at $z=q$, its local solutions are analytic and free of logarithmic monodromy; the resulting singularity at $z=q$ in the scalar equation is therefore an apparent singularity (with projective monodromy $\pm1$ coming from the square-root branch in $D$). In the geometric separation-of-variables picture~\cite{FrenkelGukovTeschner}, these apparent singularities at the zeros of $e(z)$ are the classical precursors of the extra degenerate insertions $V_{-1/(2b)}$ appearing in the Ribault--Teschner correspondence.

\section{Irregular coefficients and their semiclassical interpretation}\label{sec:semiclassical}
\subsection{Local irregular data}
Let us now examine the classical Miura dictionary in the neighborhood of an irregular puncture at $z=0$. Suppose that on the constrained Miura slice the Cartan current $\alpha(z)$ has a pole of order $r+1$:
\begin{equation}\label{eq:irregularalpha}
 \alpha(z)=\sum_{j=0}^r d_j z^{-j-1}+O(1),\qquad r\ge1,\quad d_r\ne0.
\end{equation}
Substituting this expansion into the Miura relation $t=\alpha^2-\alpha'$ determines the leading singular coefficients of $t(z)$ in the polar window $r\le n\le 2r$ solely in terms of $(d_0,\dots,d_r)$:
\begin{equation}\label{eq:taun}
 t(z)=\sum_{n=r}^{2r}\tau_n z^{-n-2}+O(z^{-r-1}),\qquad
 \tau_n=\sum_{i+j=n}d_i d_j+(n+1)d_n,
\end{equation}
where we set $d_n=0$ for $n>r$ (while regular terms in $\alpha(z)$ enter only in the subleading poles $z^{-n-2}$ with $n<r$).

For instance, for a rank-one irregular singularity ($r=1$), the two leading oper coefficients are
\begin{equation}
 \tau_2=d_1^2,\qquad \tau_1=2(d_0+1)d_1,
\end{equation}
while for rank two ($r=2$) we find
\begin{equation}
 \tau_4=d_2^2,\qquad \tau_3=2d_1d_2,\qquad
 \tau_2=d_1^2+(2d_0+3)d_2.
\end{equation}
These relations exhibit the triangular quadratic structure of the Miura map, which sends a Cartan pole of order $r+1$ to an unramified Sturm--Liouville oper with a pole of order $2r+2$.

\subsection{Comparison with Virasoro eigenvalues in the semiclassical limit}
We can now compare the classical oper coefficients \eqref{eq:taun} directly with the quantum Virasoro eigenvalues \eqref{eq:lambdan} of the irregular coherent state $|\mathbf c\rangle$. Consider the semiclassical limit $b\to\infty$ (so that $\kappa=-b^2\to-\infty$) while keeping the rescaled boson $\phi=\varphi/b$ and its current $\alpha=\partial\phi$ fixed. Since the singular part of the $\phi\phi$ OPE \eqref{eq:wakOPE} is of order $\kappa^{-1}=-b^{-2}$, $\phi$ and $\alpha$ become commuting classical fields in this limit, and the rescaled stress tensor \eqref{eq:DSquantum} converges to the Miura potential:
\begin{equation}\label{eq:classicalT}
 -\frac{T_{\mathrm{DS}}}{b^2}=-\frac{T_L}{b^2}
 =:\alpha^2:-(1+b^{-2})\partial\alpha
 \longrightarrow \alpha^2-\alpha'=(\partial\phi)^2-\partial^2\phi=t.
\end{equation}
Scaling the coherent-state parameters in \eqref{eq:coherent} proportionally to $b$ by setting
\begin{equation}
 c_j=-b d_j\quad(0\le j\le r),
\end{equation}
ensures that the singular part of $\alpha(z)=j(z)/b$ acts on $|\mathbf c\rangle$ with the finite eigenvalue $\sum_{j=0}^r d_j z^{-j-1}$, matching \eqref{eq:irregularalpha}. Dividing the exact Virasoro eigenvalues $\Lambda_n$ in \eqref{eq:lambdan} by $-b^2$ then yields, for all $r\le n\le2r$,
\begin{equation}\label{eq:matching}
 -\frac{\Lambda_n}{b^2}
 =\sum_{i+j=n}d_i d_j+(n+1)(1+b^{-2})d_n
 \xrightarrow{b\to\infty}\tau_n.
\end{equation}
Thus, in the large-$b$ semiclassical limit, the rescaled Virasoro eigenvalues of the Gaiotto--Teschner coherent state reproduce the exact leading singular coefficients $\tau_n$ ($r\le n\le 2r$) of the reduced classical oper. For the lower modes $0\le n<r$, the first-order differential operators $\sum_{s=1}^{r-n}s c_{n+s}\partial_{c_s}$ in \eqref{eq:irregularward} act nontrivially on the $e^{-b^2\mathcal F_{\mathrm{cl}}}$ semiclassical asymptotics of the conformal block, encoding the classical accessory parameters of the oper. When the irregular singularity is situated at $z=\infty$, the same analysis applies after transforming the connection and projective connection to the local coordinate $\zeta=1/z$.

\section{Lowering operators from affine screenings}\label{sec:integrals}
We now complete our picture by showing how the first-order $\beta$ field in the unreduced Wakimoto screening current \eqref{eq:affscreen} generates the spin-lowering operators that dress the scalar master function in integral solutions of the KZ equations~\cite{Wakimoto,SchechtmanVarchenko,GukovHaghighatLiuReshetikhin}.

At each finite insertion point $z_a$, let $v_a$ be a highest-weight vector satisfying $H_av_a=\ell_av_a$ and $E_av_a=0$ (either in a generic Verma module with nondegenerate contravariant form or in the finite-dimensional irreducible quotient for $\ell_a\in\mathbb Z_{\ge0}$). We realize the descendant states $F_a^r v_a$ as polynomials in a local variable $\gamma_a$ via
\begin{equation}\label{eq:screenpolynomial}
 P_a(F_a^r v_a)=(\ell_a)_{\underline r}\gamma_a^r,
 \qquad (\ell_a)_{\underline r}=\prod_{s=0}^{r-1}(\ell_a-s),
 \qquad P_a(v_a)=1.
\end{equation}
On this polynomial space, the $\slc$ generators act as first-order differential operators:
\begin{equation}\label{eq:screenpolaction}
 E_{\mathrm{pol}}=\partial_{\gamma_a},\qquad
 H_{\mathrm{pol}}=\ell_a-2\gamma_a\partial_{\gamma_a},\qquad
 F_{\mathrm{pol}}=\ell_a\gamma_a-\gamma_a^2\partial_{\gamma_a},
\end{equation}
which intertwine the algebraic relation $E_aF_a^r v_a=r(\ell_a-r+1)F_a^{r-1}v_a$. In the Wakimoto free-field realization~\eqref{eq:wakimoto}, a primary vertex operator associated with a state $u$ at $z_a$ is constructed as
\begin{equation}
 V_a[u](z)=:P_a(u)(\gamma(z))
 e^{-\ell_a\phi(z)}:,
 \qquad
 J^X(x)V_a[u](z)\sim\frac{V_a[X_a u](z)}{x-z},
\end{equation}
where $J^E=J^+$, $J^H=J^0$, and $J^F=J^-$. In particular, contracting the screening factor $\beta(w)$ with the polynomial prefactor $P_a(u)(\gamma(z_a))$ via the fundamental OPE~\eqref{eq:wakOPE} differentiates the polynomial with respect to $\gamma(z_a)$ and therefore acts on the state label $u$ by the raising operator $E_a$:
\begin{equation}\label{eq:betaraising}
 \beta(w)P_a(u)(\gamma(z_a))
 \sim\frac{(\partial_{\gamma_a}P_a(u))(\gamma(z_a))}{w-z_a}
 =\frac{P_a(E_a u)(\gamma(z_a))}{w-z_a}.
\end{equation}

To convert this linear functional on the tensor-product module into a vector-valued weight function, we equip each site with the canonical symmetric contravariant bilinear form
\begin{equation}\label{eq:screenpairing}
 \langle v_a,v_a\rangle=1,\qquad
 \langle F_a x,y\rangle=\langle x,E_a y\rangle,\qquad
 \langle H_a x,y\rangle=\langle x,H_a y\rangle,
\end{equation}
and extend it to the tensor product, writing $v_{\mathrm{top}}=\bigotimes_a v_a$ and $P=\bigotimes_a P_a$. By weight orthogonality, evaluating the polynomial at the origin extracts the pairing with the highest-weight vector: $P_a(u)(0)=\langle v_a,u\rangle$, and hence $P(u)(0)=\langle v_{\mathrm{top}},u\rangle$.

In the standard $\beta\gamma$ Fock vacuum where uncontracted $\beta$ modes annihilate the state at infinity and remaining $\gamma$ fields are evaluated at $\gamma=0$, every $\beta(w_i)$ from a screening insertion $\cS(w_i)=:\beta(w_i)e^{2\phi(w_i)}:$ contracts against one of the finite-site vertex operators $V_a[u_a](z_a)$. Stripping off the scalar bosonic factors coming from the exponentials $:e^{2\phi(w_i)}:$, a single screening insertion at $w$ produces the linear functional
\begin{equation}\label{eq:B}
 \mathcal W_w(u)=\sum_{a=1}^n\frac{\langle v_{\mathrm{top}},E_a u\rangle}{w-z_a}
 =\langle B(w)v_{\mathrm{top}},u\rangle,
 \qquad B(w)=\sum_{a=1}^n\frac{F_a}{w-z_a},
\end{equation}
where we used the contravariance property \eqref{eq:screenpairing} to move $E_a$ from the right entry $u$ to the lowering operator $F_a$ acting on $v_{\mathrm{top}}$.

Because the $\beta(w_i)\beta(w_j)$ OPE is regular and the screening currents contain no $\gamma$ fields, inserting $m$ Wakimoto screenings simply applies independent factors of this contraction. Transferring all $m$ raising operators onto $v_{\mathrm{top}}$ via the contravariant form yields the multi-screening weight functional
\begin{equation}\label{eq:screenmultiple}
 \mathcal W_{\mathbf w}(u)
 =\left\langle v_{\mathrm{top}},
 \prod_{i=1}^m\left(\sum_a\frac{E_a}{w_i-z_a}\right)u\right\rangle
 =\left\langle\prod_{i=1}^m B(w_i)v_{\mathrm{top}},u\right\rangle,
\end{equation}
where the operators $B(w_i)$ mutually commute because the lowering generators $F_a$ commute across sites and on the same site. Multiplying the vector $\prod_{i=1}^m B(w_i)v_{\mathrm{top}}$ by the scalar Coulomb-gas master function from the bosonic exponentials and integrating over $w_1,\dots,w_m$ immediately produces the Schechtman--Varchenko hypergeometric integral solutions of the regular and rank-one irregular KZ equations~\cite{SchechtmanVarchenko,GukovHaghighatLiuReshetikhin}. In the matrix-model language of Section~\ref{sec:ensembles}, expanding the product $\prod_{i=1}^m B(w_i)v_{\mathrm{top}}$ over basis states expresses each tensor component of the KZ wavefunction as a generalized resolvent expectation value inside the $\beta$-deformed eigenvalue integral.

\section{Conclusions}
In this paper we have explored the web of connections linking Liouville CFT, $\beta$-deformed matrix models, and the $\slc$ WZW model in the presence of regular and irregular singularities. Viewing these three frameworks side by side highlights the complementary roles played by their free-field building blocks:
\begin{itemize}
\item In Liouville theory and $\beta$-deformed matrix ensembles, the $V_{1/b}$ screening integrals generate scalar Coulomb-gas kernels governed by exact finite-$N$ loop equations. Packaging the eigenvalue resolvent and potential into a collective chiral boson naturally produces the background-charge improvement $Q\partial^2\varphi$ of the Virasoro stress tensor.
\item At the classical connection level, the Drinfeld--Sokolov gauge reduction of the two-component first-order operator KZ system yields a second-order scalar Sturm--Liouville oper with potential \eqref{eq:Uexpanded}. On the constrained slice $e(z)=\mu$, the derivative improvement $-\kappa h'$ in the oper potential automatically cancels the non-Cartan field $\gamma'$, leaving the purely bosonic Miura potential $t=\alpha^2-\alpha'$ whose local solutions are built from the classical screening primitive $S(z)=\int_{z_0}^z e^{2\phi(w)}\,\dd w$.
\item Near an irregular puncture of rank $r$, the rescaled Virasoro eigenvalues $\Lambda_n$ ($r\le n\le 2r$) of the Gaiotto--Teschner coherent state converge in the large-$b$ semiclassical limit directly to the leading singular coefficients $\tau_n$ of the classical Miura oper.
\item At the quantum affine level, dressing the bosonic screening exponential with the Wakimoto $\beta$ field provides the essential representation-theoretic structure needed for vector-valued WZW blocks: Wick contractions of $\beta(w_i)$ with the primary vertex polynomials act by $E_a/(w_i-z_a)$, which the contravariant bilinear form converts into the commuting spin-lowering operators $B(w_i)=\sum_a F_a/(w_i-z_a)$ of the Schechtman--Varchenko integral representation.
\end{itemize}

These results suggest several natural directions for further investigation. At the quantum representation-theoretic level, it would be very interesting to study the full BRST cohomology $H^\bullet_{\mathrm{DS}}(M)$ of general higher-rank irregular Kac--Moody modules $M$, including non-Cartan irregular data, and to track how the subleading differential Ward identities in \eqref{eq:irregularward} govern the quantum accessory parameters. Globally on higher-genus Riemann surfaces or in the presence of general ramified irregular singularities, tracking the divisor of zeros of the nilpotent current $e(z)$ and its apparent singularities will further illuminate the interplay between Drinfeld--Sokolov reduction, separation of variables~\cite{RibaultTeschner,FrenkelGukovTeschner,GaiottoLamyPoirier}, and the analytic Langlands program. Finally, the reformulation of confluent KZ integral solutions as multi-resolvent expectation values in $\beta$-deformed matrix models offers a promising arena for studying resurgent asymptotics, Stokes phenomena, and applications to two- and three-dimensional quantum gravity.

\section*{Acknowledgments}
This article grew out of a lecture series delivered at the Higher School of Economics (HSE) University, Moscow, under the auspices of the ``International Academic Cooperation'' project. I am grateful to Valery Gritsenko for the kind invitation and for suggesting the publication of these lecture notes. I would also like to thank Wei Cui, Yihua Liu, Nicolai Reshetikhin, and Fengjun Xu for valuable discussions and comments on the manuscript. The lectures and the preparation of this work were supported by the ``International Academic Cooperation'' project of HSE University and by NSFC grant 1225061018.

\appendix
\section{Mode calculation for the irregular free-field family}\label{app:modes}
From the definition of the rank-$r$ coherent state \eqref{eq:coherent}, differentiating with respect to the parameter $c_s$ brings down the creation mode $j_{-s}$:
\begin{equation}
 j_{-s}|\mathbf c\rangle=\frac s2\partial_{c_s}|\mathbf c\rangle
 \quad(1\le s\le r).
\end{equation}
When acting with the normal-ordered Virasoro generator $L_n=-\sum_p:j_{n-p}j_p:-(n+1)Qj_n$ for $n\ge0$, the terms with two nonnegative mode indices act diagonally by $-\sum_{i=0}^n c_i c_{n-i}$, while the linear background-charge term contributes $(n+1)Qc_n$. The mixed creation--annihilation terms evaluate to
\begin{equation}
 -2\sum_{s\ge1}j_{-s}j_{n+s}|\mathbf c\rangle
 =\sum_{s=1}^{r-n}s c_{n+s}\partial_{c_s}|\mathbf c\rangle,
\end{equation}
which establishes the Ward identity \eqref{eq:irregularward}. Denoting by $D_n$ the differential operator in the parameters $c_s$ appearing on the right-hand side of \eqref{eq:irregularward}, successive applications on the parameterized state satisfy $L_mL_n|\mathbf c\rangle=D_n(L_m|\mathbf c\rangle)=D_nD_m|\mathbf c\rangle$. Consequently, the operators $D_n$ obey $[D_m,D_n]=(n-m)D_{m+n}$ for $m,n\ge0$, reproducing the Virasoro algebra on the parameter space.

\section{Scalar elimination without choosing a square-root gauge}
As a direct check of the oper potential \eqref{eq:Uexpanded} without introducing the intermediate square-root gauge matrix $D$, one can write $\Psi=(x,y)^T$ in the first-order system \eqref{eq:classicalsystem} and solve the lower row algebraically for $x=(\kappa y'+hy)/e$. Substituting this expression into the upper row yields the second-order equation for $y$:
\begin{equation}
 \kappa^2y''-\kappa^2\frac{e'}e y'
 -\left(h^2+ef-\kappa h'+\kappa h\frac{e'}e\right)y=0.
\end{equation}
Eliminating the first-derivative term via the rescaling $y=e^{1/2}\chi$ immediately reproduces the Sturm--Liouville equation $(\kappa^2\partial_z^2-\cU)\chi=0$ with $\cU$ given by \eqref{eq:Uexpanded}, providing an independent verification of the sign of the $-\kappa h'$ derivative improvement.

\bibliographystyle{JHEP}
{\small\bibliography{references}}

@article{BPZ,
  author = {Belavin, A. A. and Polyakov, A. M. and Zamolodchikov, A. B.},
  title = {Infinite conformal symmetry in two-dimensional quantum field theory},
  journal = {Nuclear Physics B},
  volume = {241},
  pages = {333--380},
  year = {1984},
  doi = {10.1016/0550-3213(84)90052-X},
  url = {https://doi.org/10.1016/0550-3213(84)90052-X}
}

@article{DF,
  author = {Dotsenko, V. S. and Fateev, V. A.},
  title = {Conformal algebra and multipoint correlation functions in {2D} statistical models},
  journal = {Nuclear Physics B},
  volume = {240},
  pages = {312--348},
  year = {1984},
  doi = {10.1016/0550-3213(84)90269-4},
  url = {https://doi.org/10.1016/0550-3213(84)90269-4}
}

@article{DO,
  author = {Dorn, H. and Otto, H.-J.},
  title = {Two and three-point functions in {Liouville} theory},
  journal = {Nuclear Physics B},
  volume = {429},
  pages = {375--388},
  year = {1994},
  doi = {10.1016/0550-3213(94)00352-1},
  eprint = {hep-th/9403141},
  archivePrefix = {arXiv},
  url = {https://arxiv.org/abs/hep-th/9403141}
}

@article{ZZ,
  author = {Zamolodchikov, A. B. and Zamolodchikov, Al. B.},
  title = {Structure constants and conformal bootstrap in {Liouville} field theory},
  journal = {Nuclear Physics B},
  volume = {477},
  pages = {577--605},
  year = {1996},
  doi = {10.1016/0550-3213(96)00351-3},
  eprint = {hep-th/9506136},
  archivePrefix = {arXiv},
  url = {https://arxiv.org/abs/hep-th/9506136}
}

@article{Teschner,
  author = {Teschner, J.},
  title = {{Liouville} theory revisited},
  journal = {Classical and Quantum Gravity},
  volume = {18},
  number = {23},
  pages = {R153--R222},
  year = {2001},
  doi = {10.1088/0264-9381/18/23/201},
  eprint = {hep-th/0104158},
  archivePrefix = {arXiv},
  url = {https://arxiv.org/abs/hep-th/0104158}
}

@article{KZ,
  author = {Knizhnik, V. G. and Zamolodchikov, A. B.},
  title = {Current algebra and {Wess--Zumino} model in two dimensions},
  journal = {Nuclear Physics B},
  volume = {247},
  number = {1},
  pages = {83--103},
  year = {1984},
  doi = {10.1016/0550-3213(84)90374-2},
  url = {https://doi.org/10.1016/0550-3213(84)90374-2}
}

@article{Wakimoto,
  author = {Wakimoto, Minoru},
  title = {Fock representations of the affine {Lie} algebra {$A_1^{(1)}$}},
  journal = {Communications in Mathematical Physics},
  volume = {104},
  number = {4},
  pages = {605--609},
  year = {1986},
  url = {https://projecteuclid.org/journals/communications-in-mathematical-physics/volume-104/issue-4/Fock-representations-of-the-affine-Lie-algebra-A1_1/cmp/1104115171.full}
}

@article{FrenkelKac,
  author = {Frenkel, I. B. and Kac, V. G.},
  title = {Basic representations of affine {Lie} algebras and dual resonance models},
  journal = {Inventiones Mathematicae},
  volume = {62},
  number = {1},
  pages = {23--66},
  year = {1980},
  doi = {10.1007/BF01391662},
  url = {https://doi.org/10.1007/BF01391662}
}

@article{DS,
  author = {Drinfel'd, V. G. and Sokolov, V. V.},
  title = {{Lie} algebras and equations of {Korteweg--de Vries} type},
  journal = {Journal of Soviet Mathematics},
  volume = {30},
  number = {2},
  pages = {1975--2036},
  year = {1985},
  doi = {10.1007/BF02105860},
  url = {https://doi.org/10.1007/BF02105860}
}

@article{FeiginFrenkel,
  author = {Feigin, B. and Frenkel, E.},
  title = {Quantization of the {Drinfeld--Sokolov} reduction},
  journal = {Physics Letters B},
  volume = {246},
  number = {1-2},
  pages = {75--81},
  year = {1990},
  doi = {10.1016/0370-2693(90)91310-8},
  url = {https://doi.org/10.1016/0370-2693(90)91310-8}
}

@article{BershadskyOoguri,
  author = {Bershadsky, Michael and Ooguri, Hirosi},
  title = {Hidden {$SL(n)$} symmetry in conformal field theories},
  journal = {Communications in Mathematical Physics},
  volume = {126},
  number = {1},
  pages = {49--83},
  year = {1989},
  url = {https://projecteuclid.org/journals/communications-in-mathematical-physics/volume-126/issue-1/Hidden-rm-SLn-symmetry-in-conformal-field-theories/cmp/1104179723.full}
}

@article{Forgacs,
  author = {Forg{\'a}cs, P. and Wipf, A. and Balog, J. and Feh{\'e}r, L. and O'Raifeartaigh, L.},
  title = {{Liouville} and {Toda} theories as conformally reduced {WZNW} theories},
  journal = {Physics Letters B},
  volume = {227},
  number = {2},
  pages = {214--220},
  year = {1989},
  doi = {10.1016/S0370-2693(89)80025-5},
  url = {https://doi.org/10.1016/S0370-2693(89)80025-5}
}

@article{SchechtmanVarchenko,
  author = {Schechtman, Vadim V. and Varchenko, Alexander N.},
  title = {Arrangements of hyperplanes and {Lie} algebra homology},
  journal = {Inventiones Mathematicae},
  volume = {106},
  pages = {139--194},
  year = {1991},
  doi = {10.1007/BF01243909},
  url = {https://doi.org/10.1007/BF01243909}
}

@article{FeiginFrenkelReshetikhin,
  author = {Feigin, Boris and Frenkel, Edward and Reshetikhin, Nikolai},
  title = {{Gaudin} model, {Bethe} ansatz and critical level},
  journal = {Communications in Mathematical Physics},
  volume = {166},
  pages = {27--62},
  year = {1994},
  doi = {10.1007/BF02099300},
  eprint = {hep-th/9402022},
  archivePrefix = {arXiv},
  url = {https://arxiv.org/abs/hep-th/9402022}
}

@article{JimboNagoyaSun,
  author = {Jimbo, Michio and Nagoya, Hajime and Sun, Juanjuan},
  title = {Remarks on the confluent {KZ} equation for {$\mathfrak{sl}_2$} and quantum {Painlev{\'e}} equations},
  journal = {Journal of Physics A: Mathematical and Theoretical},
  volume = {41},
  number = {17},
  pages = {175205},
  year = {2008},
  doi = {10.1088/1751-8113/41/17/175205},
  url = {https://doi.org/10.1088/1751-8113/41/17/175205}
}

@article{NagoyaSun,
  author = {Nagoya, Hajime and Sun, Juanjuan},
  title = {Confluent primary fields in the conformal field theory},
  journal = {Journal of Physics A: Mathematical and Theoretical},
  volume = {43},
  number = {46},
  pages = {465203},
  year = {2010},
  doi = {10.1088/1751-8113/43/46/465203},
  eprint = {1002.2598},
  archivePrefix = {arXiv},
  primaryClass = {math-ph},
  url = {https://arxiv.org/abs/1002.2598}
}

@article{FFTL,
  author = {Feigin, B. and Frenkel, E. and Toledano Laredo, V.},
  title = {{Gaudin} models with irregular singularities},
  journal = {Advances in Mathematics},
  volume = {223},
  number = {3},
  pages = {873--948},
  year = {2010},
  doi = {10.1016/j.aim.2009.09.007},
  eprint = {math/0612798},
  archivePrefix = {arXiv},
  primaryClass = {math.QA},
  url = {https://arxiv.org/abs/math/0612798}
}

@article{FelderRembado,
  author = {Felder, Giovanni and Rembado, Gabriele},
  title = {Singular modules for affine {Lie} algebras, and applications to irregular {WZNW} conformal blocks},
  journal = {Selecta Mathematica},
  volume = {29},
  number = {1},
  pages = {15},
  year = {2023},
  doi = {10.1007/s00029-022-00821-y},
  eprint = {2012.14793},
  archivePrefix = {arXiv},
  primaryClass = {math.QA},
  url = {https://arxiv.org/abs/2012.14793}
}

@article{Gaiotto,
  author = {Gaiotto, Davide},
  title = {Asymptotically free {$N=2$} theories and irregular conformal blocks},
  year = {2009},
  eprint = {0908.0307},
  archivePrefix = {arXiv},
  primaryClass = {hep-th},
  note = {arXiv preprint},
  url = {https://arxiv.org/abs/0908.0307}
}

@article{GaiottoTeschner,
  author = {Gaiotto, Davide and Teschner, Joerg},
  title = {Irregular singularities in {Liouville} theory and {Argyres--Douglas} type gauge theories, {I}},
  journal = {Journal of High Energy Physics},
  volume = {12},
  pages = {050},
  year = {2012},
  doi = {10.1007/JHEP12(2012)050},
  eprint = {1203.1052},
  archivePrefix = {arXiv},
  primaryClass = {hep-th},
  url = {https://arxiv.org/abs/1203.1052}
}

@article{GaiottoLamyPoirier,
  author = {Gaiotto, Davide and Lamy-Poirier, Joel},
  title = {Irregular singularities in the {$H_3^+$} {WZW} model},
  year = {2013},
  eprint = {1301.5342},
  archivePrefix = {arXiv},
  primaryClass = {hep-th},
  note = {arXiv preprint},
  url = {https://arxiv.org/abs/1301.5342}
}

@article{RibaultTeschner,
  author = {Ribault, Sylvain and Teschner, Joerg},
  title = {{$H_3^+$}-{WZNW} correlators from {Liouville} theory},
  journal = {Journal of High Energy Physics},
  volume = {06},
  pages = {014},
  year = {2005},
  doi = {10.1088/1126-6708/2005/06/014},
  eprint = {hep-th/0502048},
  archivePrefix = {arXiv},
  url = {https://arxiv.org/abs/hep-th/0502048}
}

@article{HikidaSchomerus,
  author = {Hikida, Yasuaki and Schomerus, Volker},
  title = {{$H_3^+$} {WZNW} model from {Liouville} field theory},
  journal = {Journal of High Energy Physics},
  volume = {10},
  pages = {064},
  year = {2007},
  doi = {10.1088/1126-6708/2007/10/064},
  eprint = {0706.1030},
  archivePrefix = {arXiv},
  primaryClass = {hep-th},
  url = {https://arxiv.org/abs/0706.1030}
}

@article{FrenkelGukovTeschner,
  author = {Frenkel, Edward and Gukov, Sergei and Teschner, Joerg},
  title = {Surface operators and separation of variables},
  journal = {Journal of High Energy Physics},
  volume = {01},
  pages = {179},
  year = {2016},
  doi = {10.1007/JHEP01(2016)179},
  eprint = {1506.07508},
  archivePrefix = {arXiv},
  primaryClass = {hep-th},
  url = {https://arxiv.org/abs/1506.07508}
}

@article{AGT,
  author = {Alday, Luis F. and Gaiotto, Davide and Tachikawa, Yuji},
  title = {{Liouville} correlation functions from four-dimensional gauge theories},
  journal = {Letters in Mathematical Physics},
  volume = {91},
  pages = {167--197},
  year = {2010},
  doi = {10.1007/s11005-010-0369-5},
  eprint = {0906.3219},
  archivePrefix = {arXiv},
  primaryClass = {hep-th},
  url = {https://arxiv.org/abs/0906.3219}
}

@article{DijkgraafVafa,
  author = {Dijkgraaf, Robbert and Vafa, Cumrun},
  title = {{Toda} theories, matrix models, topological strings, and {$N=2$} gauge systems},
  year = {2009},
  eprint = {0909.2453},
  archivePrefix = {arXiv},
  primaryClass = {hep-th},
  note = {arXiv preprint},
  url = {https://arxiv.org/abs/0909.2453}
}

@article{Kostov,
  author = {Kostov, Ivan K.},
  title = {Conformal field theory techniques in random matrix models},
  year = {1999},
  eprint = {hep-th/9907060},
  archivePrefix = {arXiv},
  note = {Lecture notes, Third Claude Itzykson Meeting, Paris, 1998; arXiv preprint},
  url = {https://arxiv.org/abs/hep-th/9907060}
}

@article{ChekhovEynard,
  author = {Chekhov, Leonid and Eynard, Bertrand},
  title = {Matrix eigenvalue model: {Feynman} graph technique for all genera},
  journal = {Journal of High Energy Physics},
  volume = {12},
  pages = {026},
  year = {2006},
  doi = {10.1088/1126-6708/2006/12/026},
  eprint = {math-ph/0604014},
  archivePrefix = {arXiv},
  url = {https://arxiv.org/abs/math-ph/0604014}
}

@article{NishinakaRim,
  author = {Nishinaka, Takahiro and Rim, Chaiho},
  title = {Matrix models for irregular conformal blocks and {Argyres--Douglas} theories},
  journal = {Journal of High Energy Physics},
  volume = {10},
  pages = {138},
  year = {2012},
  doi = {10.1007/JHEP10(2012)138},
  eprint = {1207.4480},
  archivePrefix = {arXiv},
  primaryClass = {hep-th},
  url = {https://arxiv.org/abs/1207.4480}
}

@article{Rim,
  author = {Rim, Chaiho},
  title = {Irregular conformal block and its matrix model},
  year = {2012},
  eprint = {1210.7925},
  archivePrefix = {arXiv},
  primaryClass = {hep-th},
  note = {arXiv preprint},
  url = {https://arxiv.org/abs/1210.7925}
}

@article{Nagoya,
  author = {Nagoya, Hajime},
  title = {Hypergeometric solutions to {Schr{\"o}dinger} equations for the quantum {Painlev{\'e}} equations},
  journal = {Journal of Mathematical Physics},
  volume = {52},
  number = {8},
  pages = {083509},
  year = {2011},
  doi = {10.1063/1.3620412},
  eprint = {1109.1645},
  archivePrefix = {arXiv},
  primaryClass = {math-ph},
  url = {https://arxiv.org/abs/1109.1645}
}

@article{Haghighat,
  author = {Haghighat, Babak and Liu, Yihua and Reshetikhin, Nicolai},
  title = {Flat connections from irregular conformal blocks},
  year = {2023},
  eprint = {2311.07960},
  archivePrefix = {arXiv},
  primaryClass = {hep-th},
  note = {arXiv preprint},
  url = {https://arxiv.org/abs/2311.07960}
}

@article{GukovHaghighatLiuReshetikhin,
  author = {Gukov, Sergei and Haghighat, Babak and Liu, Yihua and Reshetikhin, Nicolai},
  title = {Irregular {KZ} equations and {Kac--Moody} representations},
  year = {2024},
  eprint = {2412.16929},
  archivePrefix = {arXiv},
  primaryClass = {hep-th},
  note = {arXiv preprint},
  url = {https://arxiv.org/abs/2412.16929}
}

@article{Reshetikhin,
  author = {Reshetikhin, Nicolai},
  title = {The {Knizhnik--Zamolodchikov} system as a deformation of the isomonodromy problem},
  journal = {Letters in Mathematical Physics},
  volume = {26},
  number = {3},
  pages = {167--177},
  year = {1992},
  doi = {10.1007/BF00420750},
  url = {https://doi.org/10.1007/BF00420750}
}

@article{NagoyaSunConfluent,
  author = {Nagoya, Hajime and Sun, Juanjuan},
  title = {Confluent {KZ} equations for {$\mathfrak{sl}_N$} with {Poincar{\'e}} rank 2 at infinity},
  journal = {Journal of Physics A: Mathematical and Theoretical},
  volume = {44},
  number = {28},
  pages = {285205},
  year = {2011},
  doi = {10.1088/1751-8113/44/28/285205},
  eprint = {1002.2273},
  archivePrefix = {arXiv},
  primaryClass = {math-ph},
  url = {https://arxiv.org/abs/1002.2273}
}

@article{NagoyaIrregular,
  author = {Nagoya, Hajime},
  title = {Irregular conformal blocks, with an application to the fifth and fourth {Painlev{\'e}} equations},
  journal = {Journal of Mathematical Physics},
  volume = {56},
  number = {12},
  pages = {123505},
  year = {2015},
  doi = {10.1063/1.4937760},
  eprint = {1505.02398},
  archivePrefix = {arXiv},
  primaryClass = {math-ph},
  url = {https://arxiv.org/abs/1505.02398}
}

@article{ChoiRimZhang,
  author = {Choi, Sang Kwan and Rim, Chaiho and Zhang, Hong},
  title = {{Virasoro} irregular conformal block and beta deformed random matrix model},
  journal = {Physics Letters B},
  volume = {742},
  pages = {50--54},
  year = {2015},
  doi = {10.1016/j.physletb.2015.01.003},
  eprint = {1411.4453},
  archivePrefix = {arXiv},
  primaryClass = {hep-th},
  url = {https://arxiv.org/abs/1411.4453}
}

@article{Saad:2019lba,
  author = {Saad, Phil and Shenker, Stephen H. and Stanford, Douglas},
  title = {{JT} gravity as a matrix integral},
  year = {2019},
  eprint = {1903.11115},
  archivePrefix = {arXiv},
  primaryClass = {hep-th},
  note = {arXiv preprint},
  url = {https://arxiv.org/abs/1903.11115}
}

@article{Mertens:2020hbs,
  author = {Mertens, Thomas G. and Turiaci, Gustavo J.},
  title = {{Liouville} quantum gravity -- holography, {JT} and matrices},
  journal = {Journal of High Energy Physics},
  volume = {01},
  pages = {073},
  year = {2021},
  doi = {10.1007/JHEP01(2021)073},
  eprint = {2006.07072},
  archivePrefix = {arXiv},
  primaryClass = {hep-th},
  url = {https://arxiv.org/abs/2006.07072}
}

@article{Collier:2023fwi,
  author = {Collier, Scott and Eberhardt, Lorenz and Zhang, Mengyang},
  title = {Solving 3d gravity with {Virasoro} {TQFT}},
  journal = {SciPost Physics},
  volume = {15},
  number = {4},
  pages = {151},
  year = {2023},
  doi = {10.21468/SciPostPhys.15.4.151},
  eprint = {2304.13650},
  archivePrefix = {arXiv},
  primaryClass = {hep-th},
  url = {https://arxiv.org/abs/2304.13650}
}

@article{Collier:2024mgv,
  author = {Collier, Scott and Eberhardt, Lorenz and Zhang, Mengyang},
  title = {3d gravity from {Virasoro} {TQFT}: Holography, wormholes and knots},
  journal = {SciPost Physics},
  volume = {17},
  pages = {134},
  year = {2024},
  doi = {10.21468/SciPostPhys.17.5.134},
  eprint = {2401.13900},
  archivePrefix = {arXiv},
  primaryClass = {hep-th},
  url = {https://arxiv.org/abs/2401.13900}
}

@article{Collier:2024kmo,
  author = {Collier, Scott and Eberhardt, Lorenz and M{\"u}hlmann, Beatrix and Rodriguez, Victor A.},
  title = {Complex {Liouville} String},
  journal = {Physical Review Letters},
  volume = {134},
  number = {25},
  pages = {251602},
  year = {2025},
  doi = {10.1103/k74n-s63l},
  eprint = {2409.17246},
  archivePrefix = {arXiv},
  primaryClass = {hep-th},
  url = {https://arxiv.org/abs/2409.17246}
}

@article{MarshakovMironovMorozov1991,
  author = {Marshakov, A. and Mironov, A. and Morozov, A.},
  title = {Generalized matrix models as conformal field theories. Discrete case},
  journal = {Physics Letters B},
  volume = {265},
  pages = {99--107},
  year = {1991},
  doi = {10.1016/0370-2693(91)90021-H}
}

@article{KharchevEtAl1993,
  author = {Kharchev, S. and Marshakov, A. and Mironov, A. and Morozov, A. and Pakuliak, S.},
  title = {Conformal matrix models as an alternative to conventional multi-matrix models},
  journal = {Nuclear Physics B},
  volume = {404},
  pages = {717--750},
  year = {1993},
  doi = {10.1016/0550-3213(93)90595-G},
  eprint = {hep-th/9208044},
  archivePrefix = {arXiv}
}

@article{ItoyamaMaruyoshiOota2010,
  author = {Itoyama, Hiroshi and Maruyoshi, Kazunobu and Oota, Takeshi},
  title = {The Quiver Matrix Model and {2d--4d} Conformal Connection},
  journal = {Progress of Theoretical Physics},
  volume = {123},
  pages = {957--987},
  year = {2010},
  doi = {10.1143/PTP.123.957},
  eprint = {0911.4244},
  archivePrefix = {arXiv}
}

@article{MironovMorozovShakirov2010,
  author = {Mironov, A. and Morozov, A. and Shakirov, Sh.},
  title = {Matrix Model Conjecture for Exact {BS} Periods and {Nekrasov} Functions},
  journal = {Journal of High Energy Physics},
  volume = {02},
  pages = {030},
  year = {2010},
  doi = {10.1007/JHEP02(2010)030},
  eprint = {0911.5721},
  archivePrefix = {arXiv}
}

@article{SchiappaWyllard2010,
  author = {Schiappa, Ricardo and Wyllard, Niclas},
  title = {An {$A_r$} threesome: Matrix models, {2d CFTs} and {4d $N=2$} gauge theories},
  journal = {Journal of Mathematical Physics},
  volume = {51},
  pages = {082304},
  year = {2010},
  doi = {10.1063/1.3449328},
  eprint = {0911.5337},
  archivePrefix = {arXiv}
}

@article{MarshakovMironovMorozov2009,
  author = {Marshakov, A. and Mironov, A. and Morozov, A.},
  title = {On non-conformal limit of the {AGT} relations},
  journal = {Physics Letters B},
  volume = {682},
  pages = {125--129},
  year = {2009},
  doi = {10.1016/j.physletb.2009.10.077},
  eprint = {0909.2052},
  archivePrefix = {arXiv}
}
\end{document}